\documentstyle[epsf,12pt,mycite,dina4p]{article}
\input{psfig}
\def\Pma{I\!\!P}
\def\etjet{E_T^{jet}}
\def\etajet{\eta^{jet}}

\def\etcal{E_{T,cal}^{jet}}
\def\etacal{\eta_{cal}^{jet}}
\def\phical{\varphi^{jet}_{cal}}
\def\etamax{\eta^{had}_{max}}
\def\etam{\eta_{max}}
\def\etamcal{\eta^{cal}_{max}}
\def\xg{x_\gamma^{OBS}}
\def\betao{\beta^{OBS}}

\def\etaphi{\eta-\varphi}
\def\etar{-1.5<\etajet<1}

\def\wr{134~$<W<$~277~GeV}
\def\seta{d\sigma/d\etajet}
\def\set{d\sigma/d\etjet}
\def\sw{d\sigma/dW}
\def\sxg{d\sigma/d\xg}
\def\sbeta{d\sigma/d\betao}

\def\q2{Q^2}
\def\pb1{pb$^{-1}$}
\def\gp{$\gamma p$}
\def\g2{GeV$^2$}

\newcommand{\gsim} {\mbox{\raisebox{-0.4ex}
{$\;\stackrel{>}{\scriptstyle \sim}\;$}}}
\newcommand{\lsim} {\mbox{\raisebox{-0.4ex}
{$\;\stackrel{<}{\scriptstyle \sim}\;$}}}

\begin{document}
\input epsf

\vspace{1 cm}

\begin{titlepage}

\title {
{\bf  Diffractive Dijet Cross Sections \\ in Photoproduction at HERA} \\
\author{ZEUS Collaboration } }
\date{ }

\maketitle

\vspace{4 cm}

\begin{abstract}

 Differential dijet cross sections have been measured with the ZEUS detector 
for photoproduction events in which the hadronic final state containing the 
jets is separated with respect to the outgoing proton direction by a large 
rapidity gap. The cross section has been measured as a function of the 
fraction of the photon ($\xg$) and pomeron ($\betao$) momentum participating
in the production of the dijet system. The observed $\xg$ dependence shows
evidence for the presence of a resolved- as well as a direct-photon component.
The measured cross section $\sbeta$ increases as $\betao$ increases 
indicating that there is a sizeable contribution to dijet production from
those events in which a large fraction of the pomeron momentum participates
in the hard scattering. These cross sections and the ZEUS measurements of the 
diffractive structure function can be described by calculations based on
parton densities in the pomeron which evolve according to the QCD 
evolution equations and include a substantial hard momentum component of 
gluons in the pomeron. 

\end{abstract}

\vspace{-18cm}
{\noindent
 DESY 98-045 \newline
 April 1998}
 
\setcounter{page}{0}
\thispagestyle{empty}
\pagenumbering{Roman}
\def\3{\ss}
\parindent0.cm
\parskip 3mm plus 2mm minus 2mm
 
\newpage

\begin{center}
{\Large  The ZEUS Collaboration}
\end{center}

  J.~Breitweg,
  M.~Derrick,
  D.~Krakauer,
  S.~Magill,
  D.~Mikunas,
  B.~Musgrave,
  J.~Repond,
  R.~Stanek,
  R.L.~Talaga,
  R.~Yoshida,
  H.~Zhang  \\
 {\it Argonne National Laboratory, Argonne, IL, USA}~$^{p}$
% ...

\newpage

% ...
$^{  20}$ now at ATLAS Collaboration, Univ. of Munich \\
% ...
$^{  30}$ partially supported by the Foundation for German-Russian
Collaboration DFG-RFBR \\ \hspace*{3.5mm} (grant no. 436 RUS 113/248/3 and 
no. 436 RUS 113/248/2)\\
% ...

\newpage

\begin{tabular}[h]{rp{14cm}}

$^{a}$ &  supported by the Natural Sciences and Engineering Research
          Council of Canada (NSERC)  \\
$^{b}$ &  supported by the FCAR of Qu\'ebec, Canada  \\
$^{c}$ &  supported by the German Federal Ministry for Education and
          Science, Research and Technology (BMBF), under contract
          numbers 057BN19P, 057FR19P, 057HH19P, 057HH29P \\
$^{d}$ &  supported by the MINERVA Gesellschaft f\"ur Forschung GmbH,
          the German Israeli Foundation, the U.S.-Israel Binational
          Science Foundation, and by the Israel Ministry of Science \\
$^{e}$ &  supported by the German-Israeli Foundation, the Israel Science
          Foundation, the U.S.-Israel Binational Science Foundation, and by
          the Israel Ministry of Science \\
$^{f}$ &  supported by the Italian National Institute for Nuclear Physics
          (INFN) \\
$^{g}$ &  supported by the Japanese Ministry of Education, Science and
          Culture (the Monbusho) and its grants for Scientific Research \\
$^{h}$ &  supported by the Korean Ministry of Education and Korea Science
          and Engineering Foundation  \\
$^{i}$ &  supported by the Netherlands Foundation for Research on
          Matter (FOM) \\
$^{j}$ &  supported by the Polish State Committee for Scientific
          Research, grant No.~115/E-343/SPUB/P03/002/97, 2P03B10512,
          2P03B10612, 2P03B14212, 2P03B10412 \\
$^{k}$ &  supported by the Polish State Committee for Scientific 
          Research (grant No. 2P03B08308) and Foundation for
          Polish-German Collaboration  \\
$^{l}$ &  partially supported by the German Federal Ministry for
          Education and Science, Research and Technology (BMBF)  \\
$^{m}$ &  supported by the Fund for Fundamental Research of Russian Ministry
          for Science and Edu\-cation and by the German Federal Ministry for
          Education and Science, Research and Technology (BMBF) \\
$^{n}$ &  supported by the Spanish Ministry of Education
          and Science through funds provided by CICYT \\
$^{o}$ &  supported by the Particle Physics and
          Astronomy Research Council \\
$^{p}$ &  supported by the US Department of Energy \\
$^{q}$ &  supported by the US National Science Foundation \\

\end{tabular}

\end{titlepage}
 
\newpage
\parindent 5mm
\parskip 0mm
\pagenumbering{arabic}
\setcounter{page}{1}
\normalsize
\section{\bf Introduction}

 A successful phenomenological description of the available data on soft 
diffractive processes has been obtained using Regge theory and the exchange
of a trajectory  with the quantum numbers of the vacuum, the pomeron 
($\Pma$). Hard processes in diffractive 
reactions~\cite{ingsch,berger,donlan,streng,nikzak,donlan2,cfs,soper} 
provide a tool to investigate the partonic nature of this colour-singlet
exchange and to test the universality of its properties in different
reactions.

 The first experimental evidence pointing to a partonic nature of the pomeron
was the observation of jet production in $\bar{p}p$ collisions with a 
tagged leading proton (or antiproton) made by the UA8 
Collaboration~\cite{ua8coll} following the proposal of Ingelman and 
Schlein~\cite{ingsch}. At HERA, the ZEUS and H1 Collaborations observed 
neutral-current deep-inelastic $ep$ scattering (DIS) events at high $\q2$
(where $\q2$ is the virtuality of the exchanged photon) characterised by 
a large rapidity gap from the proton direction~\cite{zelrgd93,h1lrgd94}.
The properties of these events were suggestive of a diffractive interaction
mediated by pomeron exchange between a highly-virtual photon and a proton.
The observed $\q2$ dependence indicated a pointlike nature of the 
interaction and a leading-twist mechanism. Measurements of the diffractive
structure function in DIS \cite{h1lrgd95,zelrgd95,zef2d497} were found to
be consistent within the experimental uncertainties with a diffractive
structure function which factorises into a pomeron flux factor, which depends 
on the momentum fraction lost by the proton ($x_{\Pma}$), and a pomeron 
structure function, which depends on $Q^2$ and the momentum fraction of the
struck quark within the pomeron ($\beta$). The pomeron structure function
showed an approximate scaling with $\q2$ at fixed $\beta$. More recent H1
measurements \cite{h1f2d97} have been analysed in terms of parton densities in
the pomeron which evolve according to the DGLAP equations \cite{dglap}. The
observed scaling violations indicate that most of the momentum of
the pomeron is carried by gluons.

 Diffractive photoproduction ($\q2 \approx 0$) of high transverse energy jets 
at HERA provides a process which is sensitive both to the quark (e.g. via 
$\gamma q\rightarrow qg$) and gluon (e.g. via $\gamma g \rightarrow q\bar{q}$) 
densities in the pomeron. The results on the diffractive structure 
function in DIS \cite{zelrgd95}, combined with the measured inclusive jet 
cross sections in diffractive photoproduction gave the first experimental
evidence for a gluon content of the pomeron \cite{zegpom95}. The data
indicated that between 30\% and 80\% of the momentum of the pomeron carried
by partons is due to hard gluons. The conclusion was based on the assumption
that the same pomeron flux factor (``Regge factorisation'') and the same 
pomeron parton densities apply to diffractive DIS and to diffractive jet
photoproduction at similar hard scales (``hard-scattering factorisation'').

 The CDF Collaboration has recently observed diffractive production of both
$W$-bosons \cite{wcdf} and dijet events \cite{dcdf} in $\bar{p}p$ 
collisions at $\sqrt{s}=1.8$~TeV. The analysis of these measurements yields 
an estimate of the hard-gluon content of the pomeron of $(70\pm 20)$\% 
\cite{dcdf}, which is in agreement with the result obtained at HERA 
\cite{h1f2d97,zegpom95}. However, the fraction of the total pomeron momentum
carried by partons, assuming the pomeron flux factor of Donnachie and
Landshoff \cite{donlan}, is well below our result \cite{zegpom95}.
This discrepancy could be an indication of a breakdown of hard-scattering
factorisation for diffractive hard processes in $\bar{p}p$ interactions
\cite{dcdf,juan,goul}.

\begin{figure}[h]
\begin{center}
\mbox{
\psfig{figure=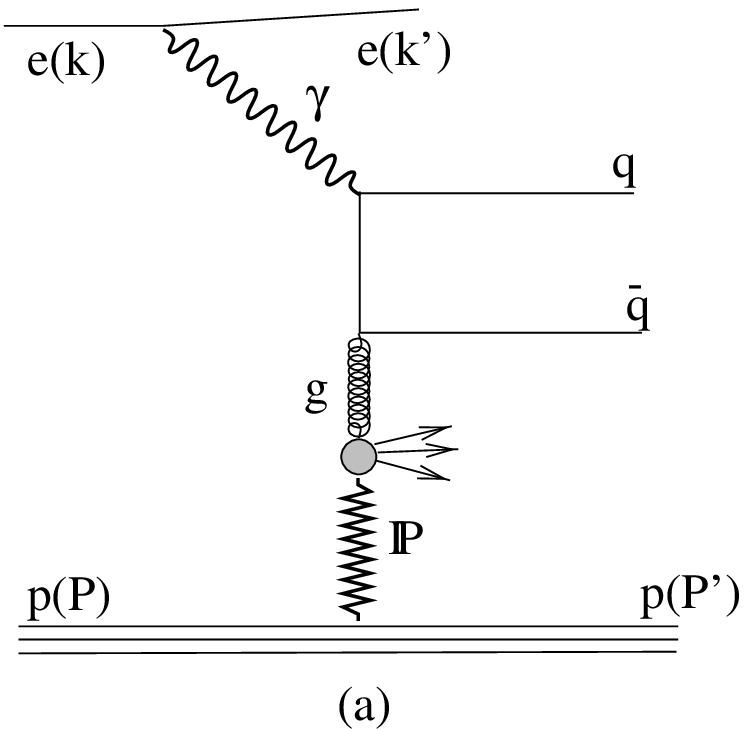,height=8.cm,width=8cm}
\hspace{1cm}
\psfig{figure=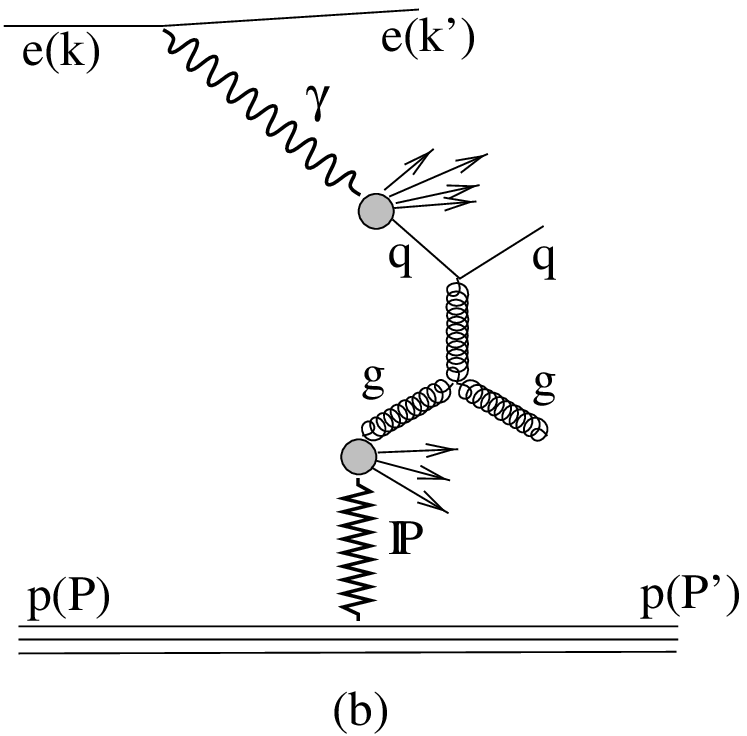,height=8.cm,width=8cm}
}
\end{center}
\caption{\label{figdra1}{ Examples of Feynman diagrams for diffractive dijet
 photoproduction: a) direct process and b) resolved process.}}
\end{figure}

  Diffractive photoproduction of dijets is sensitive to the 
underlying two-body processes. In leading-order (LO) QCD two types of
processes contribute to jet production \cite{owens,drees}: 
either the photon interacts directly with a parton in the pomeron (the 
{\it direct} process) or the photon acts as a source of partons which scatter 
off those in the pomeron (the {\it resolved} process). Examples of Feynman
diagrams for diffractive dijet photoproduction are shown in 
Figure~\ref{figdra1}. The cross section dependence on the fraction of the 
photon momentum participating in the production of the two jets, $\xg$, is 
sensitive to the presence of these two components \cite{zenov93,zedij95}. 

 In this paper, new measurements of differential cross sections for 
diffractive dijet photoproduction at HERA are presented. The results are
presented as a function of the pseudorapidity\footnote{The ZEUS coordinate 
system is defined as right-handed with the $Z$ axis pointing in the proton 
beam direction, hereafter referred to as forward, and the $X$ axis horizontal, 
pointing towards the centre of HERA. The pseudorapidity is defined as 
$\eta=-\ln(\tan\frac{\theta}{2})$, where the polar angle $\theta$ is taken 
with respect to the proton beam direction.} ($\etajet$) and transverse
energy of each jet ($\etjet$), the \gp\ centre-of-mass energy ($W$), $\xg$ 
and the fraction of the pomeron momentum ($\betao$) participating in the 
production of the two jets with highest $\etjet$ in an event. These
measurements together with those of the diffractive structure function
\cite{zelrgd95} are analysed in terms of parton densities in the pomeron
which are assumed to evolve according to the DGLAP equations. We therefore
test whether it is possible to describe both sets of data with the same
pomeron parton densities, assuming Regge factorisation.
The data sample used in this analysis was collected with the ZEUS detector
in $e^+p$ interactions at the HERA collider and corresponds to an integrated
luminosity of 2.65~\pb1.

\section{\bf Kinematics of diffractive hard processes}

 Diffractive photon-dissociative processes\footnote{These 
processes will be henceforth referred to as diffractive processes unless
otherwise stated.} in $e^+p$ collisions are characterised by a final state 
consisting of a hadronic system X, the scattered positron and the 
scattered proton
$$e^+(k) \; + \; p_i(P) \rightarrow e^+(k^{\prime})\; +\; {\rm X} \; +\;
p_f(P^{\prime}),$$
where $p_i$ ($p_f$) denotes the initial (final) state proton. The kinematics 
of this process are described in terms of four variables. Two of them describe 
the positron-photon vertex and can be taken to be the virtuality of the 
exchanged photon ($Q^2$) and the inelasticity variable $y$ defined by 
$$Q^2 = -q^2 = -(k-k^{\prime})^2 \; \; {\rm and} \; \; y = \frac{P\cdot 
q}{P\cdot k}.$$
The other two variables describe the proton vertex: the fraction of the 
momentum of the initial proton carried by the pomeron ($x_{\Pma}$), and the
square of the momentum transfer ($t$) between the initial and final
state proton, defined by
$$   x_{\Pma} = \frac{(P-P^{\prime})\cdot q}{P\cdot q} \; \; {\rm and} \; \;
     t = (P-P^{\prime})^2.$$

 The hard scale is given by $Q$ in diffractive DIS and by the jet transverse
energy in diffractive photoproduction ($Q^2 \approx 0$). In DIS, an 
additional and useful kinematic variable is given by
$$ \beta = \frac{Q^2}{2(P-P^{\prime})\cdot q} = \frac{Q^2}{s y x_{\Pma}},$$
where $\sqrt{s}$ is the $e^+p$ centre-of-mass energy. In models where the
pomeron has a partonic structure, $\beta$ is the equivalent of the Bjorken-$x$
variable in $e^+\Pma$ interactions.

 In diffractive dijet photoproduction processes the hadronic system X
contains at least two jets, 
\begin{equation}
e^+(k) \; + \; p_i(P) \rightarrow e^+(k^{\prime})\; +\; {\rm X} \; +\;
 p_f(P^{\prime}) 
\rightarrow e^+(k^{\prime})\; +\; {\rm X}({\rm jet}+{\rm jet}+{\rm X_r})\; 
+\; p_f(P^{\prime}),
 \label{reactgp}
\end{equation}
where $X_r$ denotes the remainder of the final state X not assigned to the 
two jets with the highest $\etjet$. In two-to-two massless-parton scattering,
the fractions of the photon ($x_{\gamma}$) and pomeron ($\beta_{\Pma}$) 
momenta carried by the initial-state partons (in the infinite-momentum frame 
of the parent particles) are defined by
$$ x_{\gamma} = \frac{(p_{J1}+p_{J2})\cdot q^{\prime}}{q\cdot q^{\prime}}
                   \; \; \; {\rm and} \; \; \; 
 \beta_{\Pma} = \frac{(p_{J1}+p_{J2})\cdot q}{q\cdot q^{\prime}},$$
where $p_{J1}$ and $p_{J2}$ are the momenta of the two final state partons,
$q^{\prime}$ is defined as $q^{\prime} \equiv P-P^{\prime}$ and the
approximations $q^2 \approx q^{\prime 2} \approx 0$ have been used.
For LO direct processes, $x_{\gamma}=1$. 

 Since the partons are not measurable objects, the observables $\xg$ 
\cite{zedij95} and $\betao$, defined in terms of jets, are introduced,
$$ \xg = \frac{\sum_{jets} E^{jet}_T e^{-\eta^{jet}}}{2 y E_e}
                   \; \; \; {\rm and} \; \; \; 
 \betao = \frac{\sum_{jets} E^{jet}_T e^{\eta^{jet}}}{2 x_{\Pma} E_p},$$
where $E_e$ ($E_p$) is the incident positron (proton) energy and the sum 
runs over the two jets of highest $\etjet$ in an event. The variable $\xg$
($\betao$) is an estimator of the fraction of the photon (pomeron) momentum 
participating in the production of the two jets with highest $\etjet$. The LO 
direct and resolved processes populate different regions of $\xg$, with the
direct processes being concentrated at high values of $\xg$.

 Diffractive processes give rise to a large rapidity gap between the hadronic 
system X and the scattered proton: $\Delta${\bf y}$_{GAP} = 
$~{\bf y}$_{p_f} - $~{\bf y}$^{had}_{max}$, where {\bf y}$_{p_f}$ is the
rapidity of the scattered proton and {\bf y}$^{had}_{max}$ is the rapidity
of the most-forward-going hadron belonging to the system X. Instead of 
{\bf y}$^{had}_{max}$ the pseudorapidity ($\etamax$) of the most-forward-going
hadron in the detector was used to select diffractive events.
The same signature is expected for double dissociation where the scattered
proton is replaced by a low-mass baryonic system ($N$). In this measurement,
the outgoing proton (or system $N$) was not observed.

\section{\bf A model for diffractive hard processes}

 Several models have been developed to describe diffractive hard processes 
assuming that the main mechanism for reaction~(\ref{reactgp}) is pomeron 
emission by the proton. The measurements of the diffractive structure
function \cite{zelrgd95} and of the dijet cross sections in diffractive
photoproduction presented here are analysed in terms of a model in which both
Regge and hard-scattering factorisation are assumed
\cite{ingsch,berger,donlan,streng,soper} (factorisable model). In this model,
first proposed by Ingelman and Schlein \cite{ingsch}, resolved- and
direct-photon contributions to dijet diffractive photoproduction can be
calculated. Additional coherent-pomeron contributions are expected in resolved
photoproduction from processes in which the whole pomeron initiates the hard
scattering \cite{nikzak,cfs,donlan2}. In principle such contributions would
lead to factorisation breaking. However, the inclusion of a delta-function
term at $\beta=1$ in the gluon density of the pomeron in the factorisable
model would lead to a similar $\beta$ distribution.

 In QCD, the factorisation theorem \cite{facthe} ensures that the parton 
densities (for a given hadron) extracted from measurements of DIS apply also
to other inclusive hard processes involving the same initial-state hadron. This
theorem has been recently extended to diffractive DIS \cite{collproof} which
justifies our analysis of the leading-twist diffractive structure function in 
terms of parton densities in the pomeron. This new theorem does not address the
question of Regge factorisation, which relates the properties of the pomeron in
diffractive DIS to those measured in hadron-hadron reactions.

 The proof of hard-scattering factorisation fails for diffractive hard 
processes in hadron-hadron collisions \cite{collproof}. Therefore, 
the contribution of the resolved process to diffractive dijet photoproduction 
is expected to be non-factorisable. In what follows, hard-scattering 
factorisation is assumed to hold for both processes (direct and resolved) in
the range $x_{\gamma}\gsim 0.2$. We therefore check whether a consistent set
of pomeron parton densities is able to describe simultaneously the measurements
of the diffractive structure function in DIS and the cross sections for
diffractive dijet photoproduction within the precision of the present data.

 To summarise, in this factorisable model, the pomeron is assumed to be a
source of partons which interact either with the photon (in DIS and in the
direct process of photoproduction) or with a partonic constituent of the
photon (in the resolved process of photoproduction). Calculations based on
such a model involve three basic ingredients: a) the flux of pomerons from the 
proton as a function of $x_{\Pma}$ and $t$, b) the parton densities in the
pomeron, and c) the matrix elements for the hard subprocess. For the resolved
process in photoproduction, a fourth ingredient is needed: the parton 
densities in the photon. The pomeron flux factor is extracted from 
hadron-hadron collisions using Regge phenomenology, and the 
matrix elements are computed in perturbative QCD. However, the parton 
densities are {\it a priori} unknown and have to be extracted from 
experiment.

 In what follows, the pomeron flux factor given by Donnachie and Landshoff 
(DL) \cite{donlan} has been used,
$$f_{\Pma/p}(x_{\Pma},t)=\frac{9 b_0^2}{4\pi^2} F_1(t)^2 
 x_{\Pma}^{1-2\alpha(t)} \ ,$$
where $F_1(t)$ is the elastic form factor of the proton, $b_0$ the 
pomeron-quark coupling ($b_0 = 1.8$~GeV$^{-1}$) and $\alpha(t)$ the pomeron 
trajectory ($\alpha(t) = 1.085 + 0.25 t$ with $t$ in GeV$^2$) taken from
hadron-hadron data.

 The parton densities in the pomeron, $f_{i/\Pma}(\beta,\mu^2)$ with 
$i=q,g$, depend upon the fraction $\beta$ of the pomeron momentum carried 
by parton $i$ and the scale $\mu$ at which the parton structure of the
pomeron is probed. In diffractive dijet photoproduction processes, one choice,
which is used here, is $\mu = \hat{p}_T$, where $\hat{p}_T$ is the 
transverse momentum of either of the two outgoing partons. The scale $\mu$
is set equal to $Q$ for diffractive DIS. The parton densities in the pomeron
are evolved in $\mu^2$ according to the DGLAP equations. 

 In this analysis (see Section~9), the parton densities in the 
pomeron are determined from a simultaneous fit to the ZEUS measurements of 
the diffractive structure function $\tilde{F}_2^D(\beta,Q^2)$ (see
below) \cite{zelrgd95} and the measurements of diffractive dijet cross 
sections in photoproduction presented in Section~8. 
The shapes of the resulting parton densities do not depend on the 
normalisation of the pomeron flux factor and depend only weakly on the
$x_{\Pma}$-functional form as long as this is the same in diffractive DIS and
dijet photoproduction.

\subsection{The diffractive structure function}

 For unpolarised beams, the differential cross section for diffractive
DIS can be described in terms of the diffractive structure function
$F_2^{D(4)}(\beta,Q^2,x_{\Pma},t)$
$$\frac{d^4 \sigma^{DIS}_{diff}}{d\beta dQ^2 d x_{\Pma} dt} =
 \frac{2\pi \alpha^2}{\beta Q^4} (1+(1-y)^2) 
 F_2^{D(4)}(\beta,Q^2,x_{\Pma},t),$$
where $\alpha$ is the electromagnetic coupling constant and the contribution
of the longitudinal diffractive structure function is neglected. An 
integration over the entire range of $t$ defines the diffractive
structure function $F_2^{D(3)}(\beta,Q^2,x_{\Pma})$ \cite{ingpry}
as measured in \cite{zelrgd95}:
$$\frac{d^3 \sigma^{DIS}_{diff}}{d\beta dQ^2 d x_{\Pma}} =
 \frac{2\pi \alpha^2}{\beta Q^4} (1+(1-y)^2) 
 F_2^{D(3)}(\beta,Q^2,x_{\Pma}).$$
To illustrate the $\beta$ and $Q^2$ dependence of 
$F_2^{D(3)}(\beta,Q^2,x_{\Pma})$, the structure function
$\tilde{F}_2^D(\beta,Q^2)$ was also measured in \cite{zelrgd95}:
$$\tilde{F}_2^D(\beta,Q^2)\equiv\int_{x_{\Pma{\rm min}}}^{x_{\Pma {\rm max}}}
 dx_{\Pma} F_2^{D(3)}(\beta,Q^2,x_{\Pma}).$$

 In the factorisable model, the diffractive structure function $F_2^{D(4)}$
is assumed to be of the form
$$ F_2^{D(4)}(\beta,Q^2,x_{\Pma},t) = f_{\Pma/p}(x_{\Pma},t) \cdot 
 F_2^{\Pma}(\beta,Q^2),$$
where at LO the pomeron structure function $F_2^{\Pma}$ depends on the
pomeron parton densities as given by
$$ F_2^{\Pma}(\beta,Q^2) = \sum_{j} e_j^2 \beta f_{q_j/\Pma}(\beta,Q^2).$$
In this expression, $e_j$ is the electric charge of quark $q_j$ and the
sum runs over all quark flavours which contribute at the given value 
of $Q^2$. Therefore, the LO calculation for $\tilde{F}_2^D$ in this
model is of the form
$$\tilde{F}_2^D(\beta,Q^2) = \sum_{j} e_j^2 \beta f_{q_j/\Pma}(\beta,Q^2)
\cdot \int_{x_{\Pma{\rm min}}}^{x_{\Pma {\rm max}}} dx_{\Pma} \ dt \
f_{\Pma/p}(x_{\Pma},t).$$

 At next-to-leading order (NLO), $F_2^{\Pma}$ also depends on the
gluon density in the pomeron. The measurements of the diffractive
structure function are analysed (see Section~9) in terms of NLO
QCD calculations.

\subsection{Dijet cross sections in diffractive photoproduction}

 The dijet cross sections in diffractive photoproduction contain
contributions from both the direct and resolved processes. As an example, the
contribution of the direct process to the cross section for
reaction~(\ref{reactgp}) is given by
$$ \sigma_{dir} = \int dy f_{\gamma/e}(y) \int\int dx_{\Pma} dt
  f_{\Pma/p}(x_{\Pma},t) \sum_i \int d\beta
  \sum_{j,k} \int d\hat{p}^2_T
  \frac{d\hat{\sigma}_{i+\gamma \rightarrow j + k }}{
  d\hat{p}^2_T}(\hat{s},\hat{p}^2_T,\mu^2) \; \cdot \;
  f_{i/\Pma}(\beta,\mu^2), $$
where $f_{\gamma/e}$ is the flux of photons from the
positron\footnote{ The $Q^2$ dependence has been integrated out using
the Weizs\"{a}cker-Williams approximation.}. The sum in $i$ runs over 
all possible types of partons present in the pomeron. The sum in $j$ and $k$
runs over all possible types of final state partons and
$\hat{\sigma}_{i+\gamma \rightarrow j+k}$ is the cross section for the
two-body collision $i + \gamma \rightarrow j + k$ and depends on
the square of the centre-of-mass energy ($\hat{s}$), the transverse
momentum of the two outgoing partons ($\hat{p}_T$) and the momentum
scale ($\mu$) at which the strong coupling constant ($\alpha_s(\mu^2)$)
is evaluated.

\section{Experimental conditions}

 During 1994 HERA operated with protons of energy $E_p=820$~GeV and positrons 
of energy $E_e=27.5$~GeV. The ZEUS detector is described in detail in 
\cite{sigtot,status}. The main subdetectors used in the present analysis are 
the central tracking system positioned in a 1.43~T solenoidal magnetic field 
and the uranium-scintillator sampling calorimeter (CAL). The tracking system 
was used to establish an interaction vertex and to cross-check the energy 
scale of the CAL. Energy deposits in the CAL were used to find jets and 
to measure jet energies. The CAL is hermetic and consists of 5918~cells
each read out by two photomultiplier tubes. Under test beam conditions, the CAL
has energy resolutions of 18\%/$\sqrt{E}$ for electrons and 35\%/$\sqrt{E}$
for hadrons. Jet energies are corrected for the energy lost in inactive
material in front of the CAL. This material is typically about one radiation 
length. The effects of uranium noise were minimised by discarding cells in the
inner (electromagnetic) or outer (hadronic) sections if they had energy
deposits of less than 60~MeV or 110~MeV, respectively. The luminosity was
measured from the rate of the brems\-strahlung process
$e^+ p \rightarrow e^+ p \gamma$. A three-level trigger was used to select 
events online \cite{status}. At the third level, where the full event
information is available, the events were required to have at least two jets 
with jet transverse energy in excess of 3.5~GeV and jet pseudorapidity below
2.0 reconstructed using the CAL cell energies and positions as input to a 
cone algorithm (see Section~6).

\section{Monte Carlo simulation}

 The response of the detector to jets and the correction factors for the cross 
sections for dijet production with a large rapidity gap were determined from 
Monte Carlo samples of events.

 The program PYTHIA~5.7 \cite{pythia} was used to generate standard 
(non-diffractive) hard photoproduction events for resolved and direct 
processes. The photon momentum spectrum was calculated using the
Weizs\"{a}cker-Williams approximation. Events were generated using GRV-HO 
\cite{grv} for the photon parton distributions and MRSA \cite{mrsa} for the 
proton parton distributions. The partonic processes were simulated using LO 
matrix elements, with the inclusion of initial- and final-state parton showers.
Fragmentation into hadrons was performed using the LUND string model 
\cite{lund} as implemented in JETSET \cite{jetset}. Samples of events were 
generated with different values of the cutoff on the transverse momentum of 
the two outgoing partons, starting at $\hat p_{Tmin}= 2.5$~GeV.

 Diffractive processes were simulated using the program POMPYT~2.5\footnote{This
version of POMPYT has been modified to make use of pomeron parton
densities which evolve with the scale according to the DGLAP equations.}
\cite{pompyt}. This is a Monte Carlo program where, within the framework 
provided by PYTHIA, the proton emits a pomeron whose partonic constituents 
subsequently take part in a hard scattering process with the photon or its 
constituents. For the resolved processes, the parton densities of the photon 
were parametrised according to GS-HO \cite{gs}\footnote{The correction 
factors depend very weakly on the specific set of photon parton distributions 
used.} evaluated at $\hat{p}_T$. The parton densities in the pomeron were
parametrised according to a hard distribution $\beta (1-\beta)$ and the DL
form was used for the pomeron flux factor.

 All generated events were passed through the ZEUS detector and trigger
simulation programs \cite{status}. They were reconstructed and analysed by
the same program chain as the data. The resulting Monte Carlo distributions
agree reasonably well with the data.

\section{Jet search and reconstruction of kinematic variables}

 An iterative cone algorithm in the $\etaphi$ plane is used to reconstruct 
jets from the energy measured in the CAL cells for both data and simulated 
events, and also from the final-state hadrons for simulated events. A detailed
description of the algorithm can be found in \cite{zeoct97}. The jets 
reconstructed from the CAL cell energies are called $cal$ jets and the 
variables associated with them are denoted by $\etcal$, $\etacal$ and 
$\phical$. The axis of the jet is defined according to the Snowmass convention
\cite{snow}, where $\etacal$ ($\phical$) is the transverse-energy weighted
mean pseudorapidity (azimuth) of all the CAL cells belonging to that jet.
The cone radius used in the jet search was set equal to $1$. 

 For the Monte Carlo events, the same jet algorithm is also applied to the 
final-state particles. The jets found are called $hadron$ jets and
the variables associated with them are denoted by $E^{jet}_{T,had}$, 
$\eta^{jet}_{had}$, and $\varphi^{jet}_{had}$. $Hadron$ jets with
$E^{jet}_{T,had}>6$ GeV and $-1.5<\eta^{jet}_{had}<1$ are selected.

 The comparison of the reconstructed jet variables between the $hadron$ and 
the $cal$ jets in simulated events~\cite{zeoct94} shows no significant 
systematic shift in the angular variables $\etacal$ and $\phical$ with 
respect to $\eta^{jet}_{had}$ and $\varphi^{jet}_{had}$. The resolutions
are 0.07~units in $\etacal$ and $5^{\circ}$ in $\phical$. The 
transverse energy of the $cal$ jet underestimates that of the $hadron$ jet by 
an average amount of 16\% with an r.m.s. of 11\%. The transverse 
energy corrections to $cal$ jets averaged over the azimuthal angle were 
determined using the Monte Carlo samples of events. These 
corrections are constructed as multiplicative factors, $C(\etcal,\etacal)$, 
which, when applied to the $E_T$ of the $cal$ jets, give the true
transverse energies of the jets, $\etjet=C(\etcal,\etacal) \times \etcal$
\cite{zeoct94}. These corrections mainly take into account the 
energy losses due to the inactive material in front of the CAL. 

 The \gp\ centre-of-mass energy $W=\sqrt{y s}$ is estimated using the 
method of Jacquet-Blondel~\cite{jacblo},
\begin{equation}
W_{JB} = \sqrt{2 E_p (E-P_Z)}, 
\end{equation}
where $E$ is the total energy as measured by the CAL, $E=\sum_i E_i$, and 
$P_Z$ is the $Z$-component of the vector $\vec{P}=\sum_i E_i \vec r_i$; in both
cases the sum runs over all CAL cells, $E_i$ is the energy of the calorimeter
cell $i$ and $\vec r_i$ is a unit vector along the line joining the
reconstructed vertex and the geometric centre of the cell $i$.
Due to the energy lost in the inactive material in front of the CAL and to
particles lost in the rear beampipe, $W_{JB}$ systematically underestimates
the true $W$ by approximately 13\%, an effect which is adequately reproduced
in the Monte Carlo simulation of the detector. Monte Carlo studies show a
resolution of 6\% for $W_{JB}$ in the range $120 < W_{JB} < 251$~GeV. The
reconstructed value $W_{JB}$ is corrected to the true $W$ by means of the
Monte Carlo samples of events.

 The variables of the dijet system $\xg$ and $\betao$ are reconstructed as
\begin{eqnarray}
x^{OBS}_{\gamma,cal}
&=&2 E_p \frac{\sum_{jets} \etcal e^{-\etacal}}{W^2_{JB}}, 
\; \; {\rm and} \\
\beta^{OBS}_{cal}
&=&\frac{\sum_{jets} \etcal e^{\etacal}}{2 x^{cal}_{\Pma} E_p},
\end{eqnarray}
where $x^{cal}_{\Pma}$ is determined from the energies and angles measured 
in the CAL using
\begin{equation}
 x^{cal}_{\Pma} = \frac{E^2 - P_X^2 - P_Y^2 - P_Z^2}{W^2_{JB}},
\end{equation}
where $P_X$ ($P_Y$) is the $X$($Y$)-component of the vector $\vec{P}$.
Note that in the formulae for $\xg$ and $\betao$
many systematic uncertainties in the measurement of energy by the CAL cancel
out. There are no significant systematic shifts in the variables 
$x^{OBS}_{\gamma,cal}$ and $\beta^{OBS}_{cal}$ with respect to $\xg$ and 
$\betao$. Therefore, no corrections are needed for these variables
($\xg\approx x^{OBS}_{\gamma,cal}$ and $\betao\approx\beta^{OBS}_{cal}$).
The resolution in $\xg$ ($\betao$) is 0.06~units in the region
$\xg > 0.2$ ($\betao > 0.4$).

 In the analysis presented here, large-rapidity-gap events are selected by 
using the variable $\etamcal$, which is defined as the pseudorapidity of the 
most-forward CAL cluster exceeding 400~MeV~\cite{zelrgd93}. The $\etam$ 
variable for the Monte Carlo events at the CAL level is defined in the same 
way as in the data ($\etamcal$). At the hadron level, $\etamax$ is
calculated as the pseudorapidity of the most-forward particle with energy in 
excess of 400~MeV and pseudorapidity below~4.5~\cite{zegpom95}. From Monte Carlo
studies the resolution on $\etamcal$ is 0.1~units for $\etamcal < 1.8$.

\section{\bf Data selection}

 Events from quasi-real photon-proton collisions were selected offline using 
criteria similar to those reported previously \cite{zegpom95} and briefly
discussed here. A search for jet structure using the CAL cells is performed
and events with at least two jets of $\etjet>6$~GeV and $\etar$ are retained.
The contamination from beam-gas interactions, cosmic showers and beam-halo 
muons is negligible after imposing a cut on the vertex reconstructed from
three or more tracks. Neutral current DIS events are removed from the sample 
by identifying the scattered positron candidate using the pattern of energy 
distribution in the CAL~\cite{zenov93}. 

 Large-rapidity-gap events are selected by using the variable $\etamcal$ and 
events with $\etamcal < 1.8$ are kept for further analysis~\cite{zegpom95}.
The selected sample consists of events from $e^+ p$ interactions with 
$\q2\lsim 4$~\g2\ and a median of $\q2\approx 10^{-3}$~\g2 . The event
sample is then restricted to the kinematic range \wr\ using the corrected
value of $W_{JB}$. The data sample thus obtained consists of 403~events and
represents 1.3\% of the whole dijet sample in the same kinematic region
except for the requirement on $\etamcal$. The range in $x_{\Pma}$ spanned
by the data is from $0.001$ to $0.03$ with a median of $x_{\Pma}\approx 0.009$.

\section{\bf Dijet cross sections}

 Using the selected data sample of dijet events, differential dijet cross 
sections have been measured in the above kinematic region with
the most-forward-going hadron at $\etam < 1.8$. The cross sections have
been measured as a function of $\etajet$, $\etjet$, $W$, $\xg$ and
$\betao$ for dijet production with $\etjet >6$~GeV and $\etar$. For
each cross section, an integration over the remaining variables is implied.
The cross sections refer to jets at the hadron level with a cone radius of
$1$ unit in the $\etaphi$ plane. 

 The Monte Carlo samples of events generated using POMPYT were used to compute
acceptance corrections to the $\etajet$, $\etjet$, $W$, $\xg$ and $\betao$ 
distributions. These corrections take into account the efficiency of the 
trigger, the selection criteria and, the purity and efficiency of the 
reconstruction of jets. They also correct for the migrations in the variable 
$\etamcal$ and yield cross sections for the true rapidity gap determined by 
$\etamax$. The cross sections are obtained by applying bin-by-bin corrections 
to the distributions of the data.

\subsection{\bf Background and systematic uncertainties of the
measurements}

 The contribution from non-diffractive processes has been estimated using 
a sample of direct and resolved processes generated with the PYTHIA Monte 
Carlo\footnote{These calculations give a good description of the inclusive 
jet differential cross sections (without the large-rapidity-gap requirement) 
in the range $-1<\etajet<1$ \cite{zeoct94}.}. The fraction of 
large-rapidity-gap events in PYTHIA is strongly suppressed, although 
fluctuations in the final-state system may give rise to a rapidity gap in the 
forward region in a small fraction of events. The contribution from
non-diffractive processes as modelled by PYTHIA is approximately 20\%
in $\sw$. This contribution, which has been subtracted bin-by-bin from the 
data, is listed in Table~\ref{tabsec1}.

 Events with diffractively dissociated protons with masses $M_N$ less than 
$\sim 4$~GeV also contribute to the data set \cite{zelrgd95}. Since the 
pomeron flux factor of Donnachie and Landshoff accounts only for processes
in which the proton remains intact after the collision and as the measurements
are based upon a large-rapidity-gap requirement, the contribution to the
measured cross sections from double dissociation has to be taken
into account when comparing to model predictions. A comparison of the
measurements of the diffractive structure function in DIS based on the
detection of the final-state scattered proton~\cite{zef2d497} and on the
$M_X$-method~\cite{zemx97eps}, shows that double dissociation with 
$M_N\lsim$~4~GeV contributes $(31\pm 13)\%$ to the cross section. This
estimation has been cross-checked with a study of exclusive
vector-meson production~\cite{VMN639eps} yielding consistent results.
Assuming Regge factorisation, the same contribution is expected to be present 
in the measurements of diffractive dijet photoproduction. Instead of
subtracting the measurements for this contribution, the predictions of the
MC program POMPYT (see next section) have been scaled up by the appropriate
factor. Since this contribution affects equally the measurements of dijet
cross sections and those of the diffractive structure function, the results
of the QCD analysis (except for the overall normalisation) do not depend on
its exact value.

 The statistical errors are indicated as the inner error bars in all the 
figures presented below. A detailed study of the sources contributing to the 
systematic uncertainties of the measurements was carried out \cite{juanpu}. 
These uncertainties are classified into six groups:
\begin{itemize}
 \item The $\etamcal$ variable in the data and simulated events was
      recomputed after removing the CAL cells with $\eta > 3.25$ in order to
      check the dependence on the detailed simulation of the forward region
      of the detector. This resulted in changes within $\pm 10$\%.
 \item The energy threshold in the computation of $\etamcal$ for data and
      simulated events was varied by $\pm 100$~MeV,
      yielding changes within $\pm 10$\%.
 \item The amount of the non-diffractive contribution subtracted from the data 
      was varied by $\pm 30$\%, yielding changes within $\pm 15$\%.
 \item The relative contribution of quarks and gluons in the pomeron used in 
      the simulated events was varied within the range obtained by
      the QCD analysis presented in the next section. The resulting
      changes are below 15\% except for $\sxg$ in the lowest measured 
      $\xg$-point (+40\%) and $\sbeta$ in the highest measured 
      $\betao$-point (-20\%).
 \item Variations in the simulation of the trigger and a variation of the
      cuts used to select the data within the ranges allowed by the comparison
      between data and Monte Carlo simulations yielded negligible changes.
\end{itemize}
All these systematic uncertainties have been added in quadrature to the 
statistical errors and are shown as the outer error bars in the figures. 
\begin{itemize}
 \item The absolute energy scale of the $cal$ jets in the simulated events 
      was varied by $\pm 5$\% \cite{dij97}, resulting in changes
      of approximately $\pm 25$\%. This uncertainty is the largest
      source of systematic error and is highly correlated between measurements
      at different points. It is shown as a shaded band in each figure.
\end{itemize}
In addition, there is an overall normalisation uncertainty of 1.5\% from the 
luminosity determination which is not included. The results are presented 
in Table~\ref{tabsec1}.

\subsection{\bf Results}

\begin{figure}
\epsfysize=18cm
\centerline{\epsffile{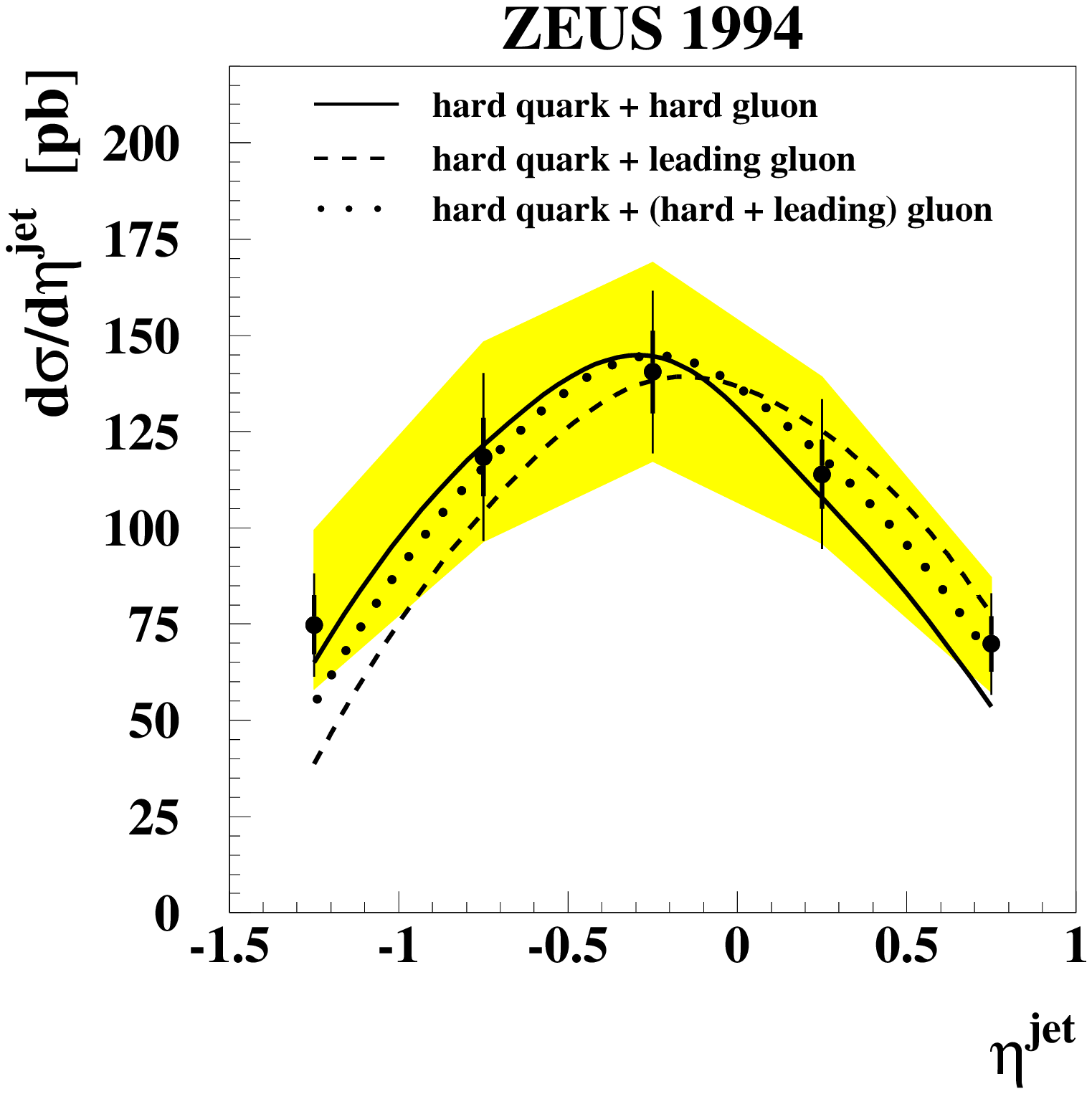}}
\vspace{-1.5cm}
\caption{\label{figdra2}{ Measured jet cross section $\seta$ in dijet events
 (see text) integrated over $\etjet>6$~GeV in the kinematic region defined 
 by $\q2\leq 4$ \g2\ and \wr\ with the most-forward-going hadron at 
 $\etam < 1.8$ (black dots). The contribution from non-diffractive 
 processes (see Table~\ref{tabsec1}) has been subtracted. The measurements
 contain an estimated $(31\pm 13)\%$ contribution from double
 dissociation. The inner error bars represent the statistical
 errors of the data, and the outer error bars show the statistical and
 systematic uncertainties $-$not associated with the absolute energy scale of
 the jets$-$ added in quadrature. The shaded band displays the uncertainty
 due to the absolute energy scale of the jets. For comparison, the results
 of the QCD fits, which have been scaled up to account for the
 contribution from double dissociation, are shown (see text). The results
 of the QCD fits have been obtained by an integration over the same bins
 as for the data and are presented as smooth curves joining the calculated
 points.}}
\end{figure}

\begin{figure}
\epsfysize=18cm
\centerline{\epsffile{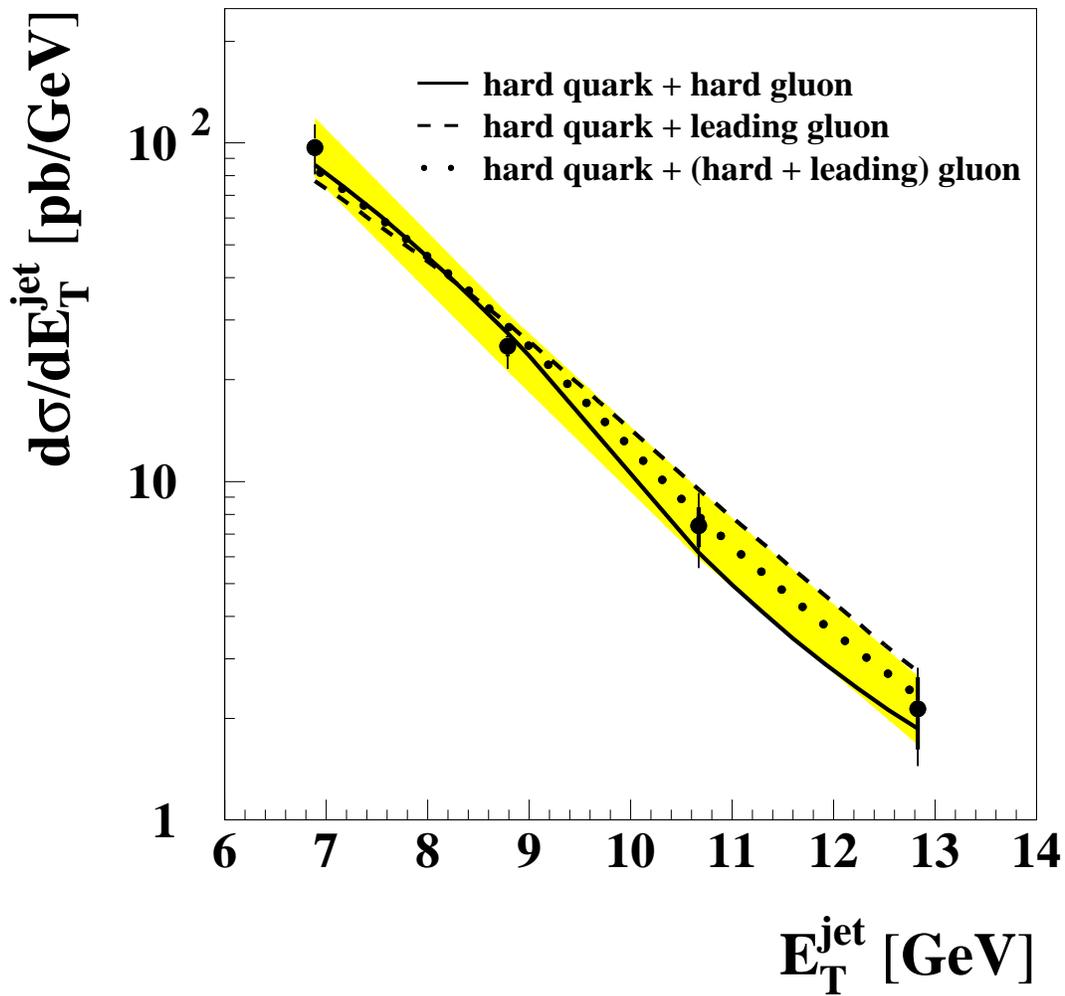}}
\vspace{-1.5cm}
\caption{\label{figdra3}{ Measured jet cross section $\set$ in dijet events
 (see text) integrated over $\etar$. Other details as in Figure~\ref{figdra2}.}}
\end{figure}

 The cross section $\seta$, shown in Figure~\ref{figdra2}, for the selected
region of phase space has been measured in the $\etajet$ range between 
$-1.5$ and 1 integrated over $\etjet> 6$~GeV. The measured cross section
$\seta$ shows a broad central maximum in $\etajet$.
The cross section $\set$ measured in the range of $\etjet$ between
$6$ and $14$~GeV and integrated over $\etar$ is presented in 
Figure~\ref{figdra3}. The cross section $\set$ shows a steep fall-off as a
function of $\etjet$. Note that in the data of Figures~\ref{figdra2} and
\ref{figdra3} each of the two jets with highest $\etjet$ in an event
contributes to the cross section. 

\begin{figure}
\epsfysize=18cm
\centerline{\epsffile{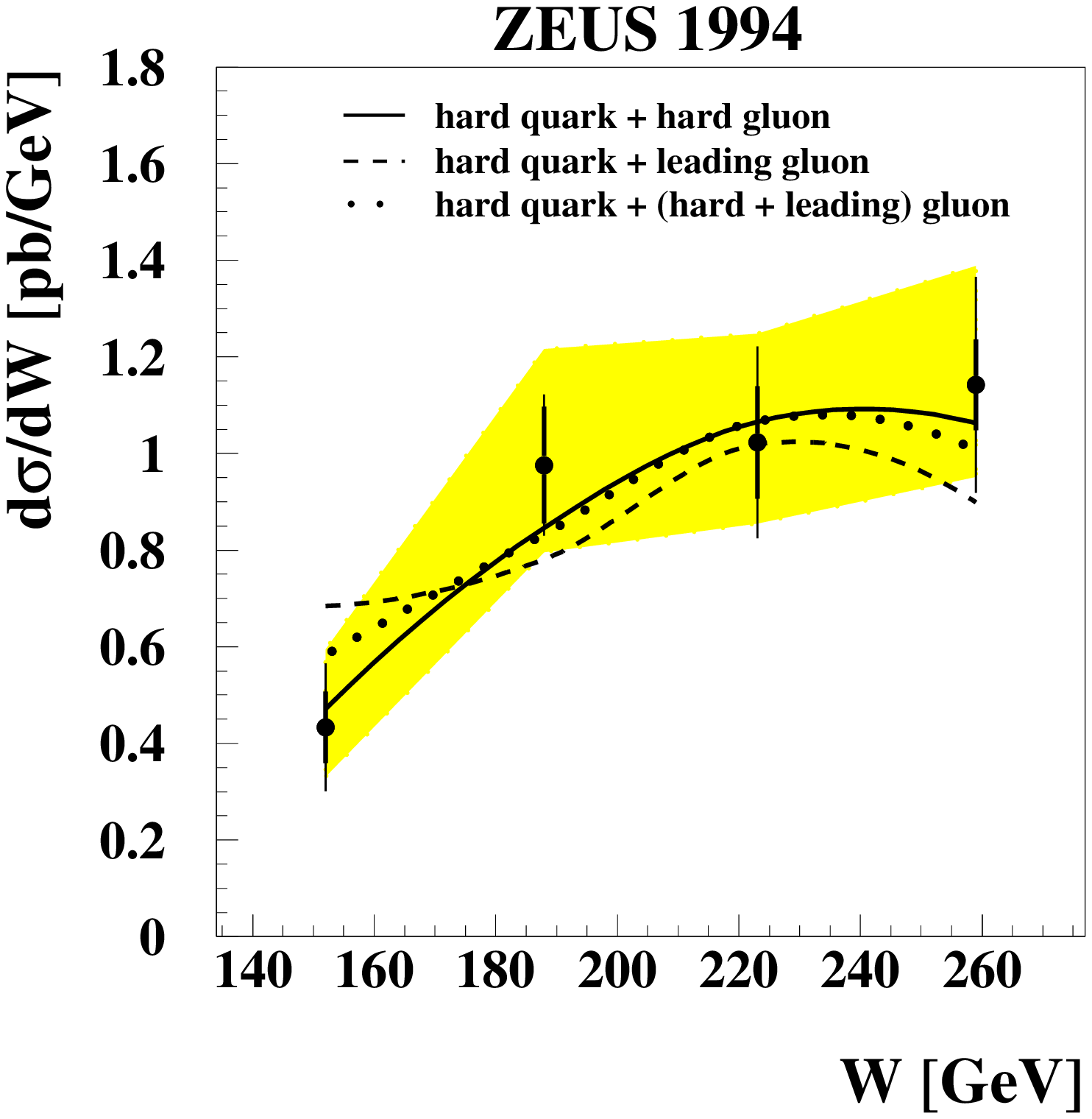}}
\vspace{-1.5cm}
\caption{\label{figdra4}{ Measured dijet cross section $\sw$ integrated
 over $\etjet >6$~GeV and $\etar$. Other details as in Figure~\ref{figdra2}.}}
\end{figure}

 The cross section $\sw$ for the selected region of phase space has been 
measured in the $W$ range between 134~GeV and 277~GeV. The cross section, 
shown in Figure~\ref{figdra4}, falls for low values of $W$
but remains fairly constant for large values of $W$.

 The cross section $\sxg$ has been measured in the $\xg$ range between
$0.2$ and $1$. The cross section $\sxg$, shown in Figure~\ref{figdra5},
peaks at high values of $\xg$ with a pronounced tail that extends to 
low-$\xg$ values. This result shows the presence of both a resolved- 
(low-$\xg$) and a direct-photon (high-$\xg$) component in diffractive dijet 
photoproduction.

 The cross section $\sbeta$, measured in the $\betao$ range between $0.4$ 
and $1$, is shown in Figure~\ref{figdra6}. The cross section increases
as $\betao$ increases and shows that there is a sizeable contribution to
dijet production from those events in which a large fraction of the pomeron
momentum participates in the hard scattering. 

 The results are compared with the predictions of the MC program POMPYT using
various parton distributions which are described in the following section.

\section{\bf QCD analysis}

 A combined QCD analysis of the ZEUS measurements of the diffractive
structure function $\tilde{F}_2^D(\beta,Q^2)$ in DIS \cite{zelrgd95}
and of the measured dijet cross sections in diffractive photoproduction 
presented in the previous section has been performed following the proposal 
by Collins et al.~\cite{collins1}. This procedure assumes both 
hard-scattering factorisation and Regge factorisation of pomeron exchange in 
diffractive DIS and diffractive dijet photoproduction.

 The diffractive structure function $\tilde{F}_2^D(\beta,Q^2)$ was
measured \cite{zelrgd95} for $Q^2$ values between 10 and 63~GeV$^2$, and
$\beta$ values between 0.175 and 0.65, by means of
an integration over the measured range of $x_{\Pma}$, 
$6.3 \cdot 10^{-4} < x_{\Pma} < 10^{-2}$. The fits to the DIS data are
based on full NLO QCD calculations. On the other hand, the
fits to the photoproduction data use calculations with LO matrix elements plus 
parton-shower as incorporated in POMPYT. Both direct and resolved processes
are included.

 In the calculations of the diffractive dijet cross sections using POMPYT, 
$\alpha_s(\mu^2)$  and the parton densities in the pomeron and the photon
are evaluated at $\mu^2=\hat{p}_T^2$. These computations may be affected by 
higher-order QCD corrections, which are expected to mainly change the 
normalisation (i.e. generate a $K$-factor). The agreement found between the
PYTHIA calculations of the inclusive jet differential cross sections and the
ZEUS measurements \cite{zeoct94} indicates that in the case of the
non-diffractive contribution the $K$-factor is close to unity, within an
uncertainty of $\pm 30$\%. The $K$-factor in the case of POMPYT is expected to
be similar (with a similar uncertainty), as the same hard subprocesses are
involved in the calculation of the jet cross sections.

 Each of the fits is represented by a parametrisation of the initial 
distributions at $\mu^2_0 = 4$~\g2\ for the quarks 
($f_{u/\Pma}(\beta,\mu^2)=f_{\bar{u}/\Pma}=f_{d/\Pma}=f_{\bar{d}/\Pma}$)
and for the gluon ($f_{g/\Pma}$). The other 
quark distributions are assumed to be zero at the initial scale. The parton 
distributions are evolved in $\mu^2$ according to the DGLAP equations at NLO 
with the number of flavours set equal to five, the $\overline{MS}$-scheme 
with $\Lambda^{(5)}_{\overline{MS}}=152$~MeV and using the program from
the CTEQ group \cite{cteq}.

\begin{figure}
\epsfysize=18cm
\centerline{\epsffile{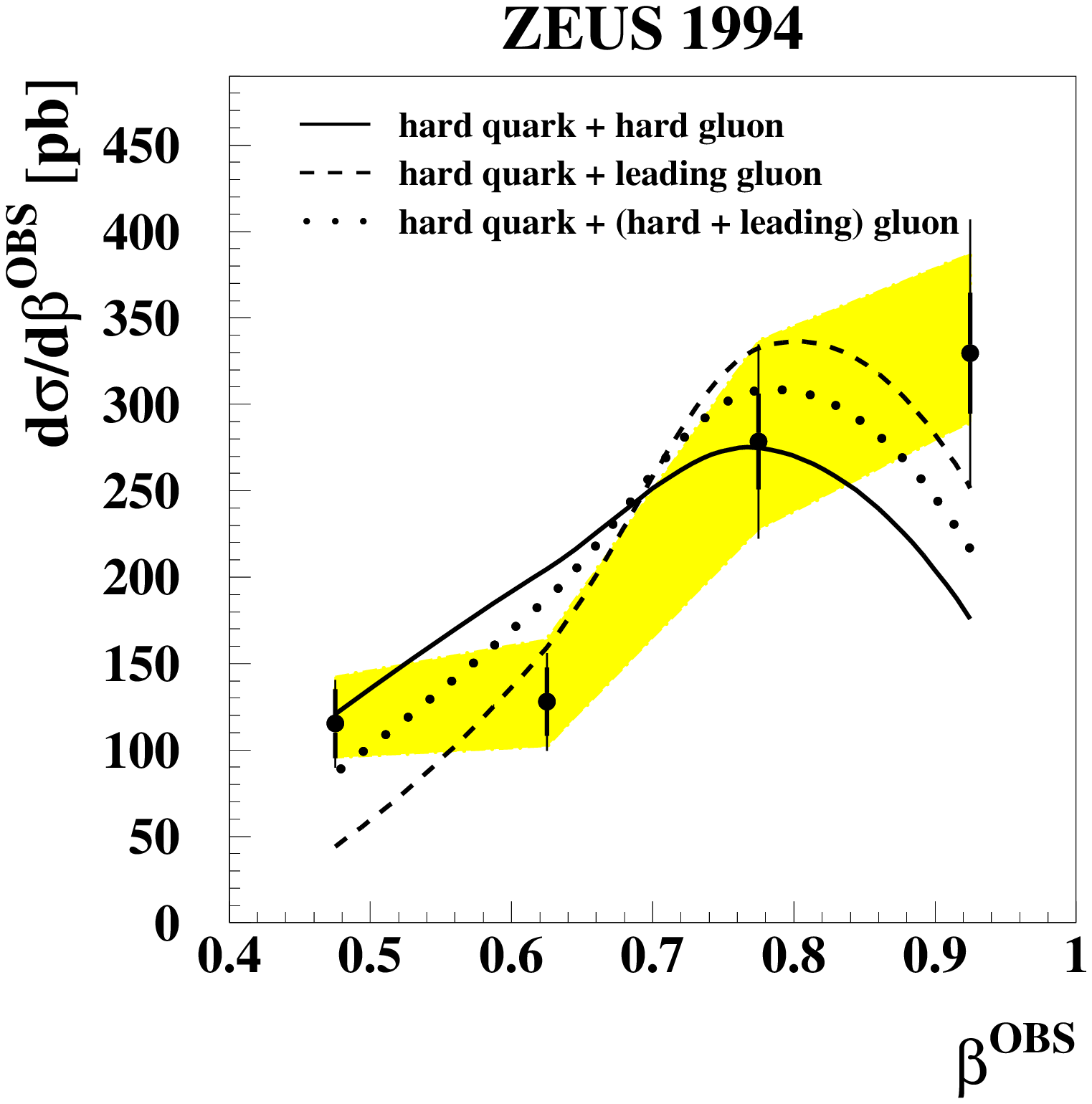}}
\vspace{-1.5cm}
\caption{\label{figdra6}{ Measured dijet cross section $\sbeta$ integrated 
 over $\etjet >6$~GeV and $\etar$. Other details as in Figure~\ref{figdra2}.}}
\end{figure}

\begin{figure}
\epsfysize=18cm
\centerline{\epsffile{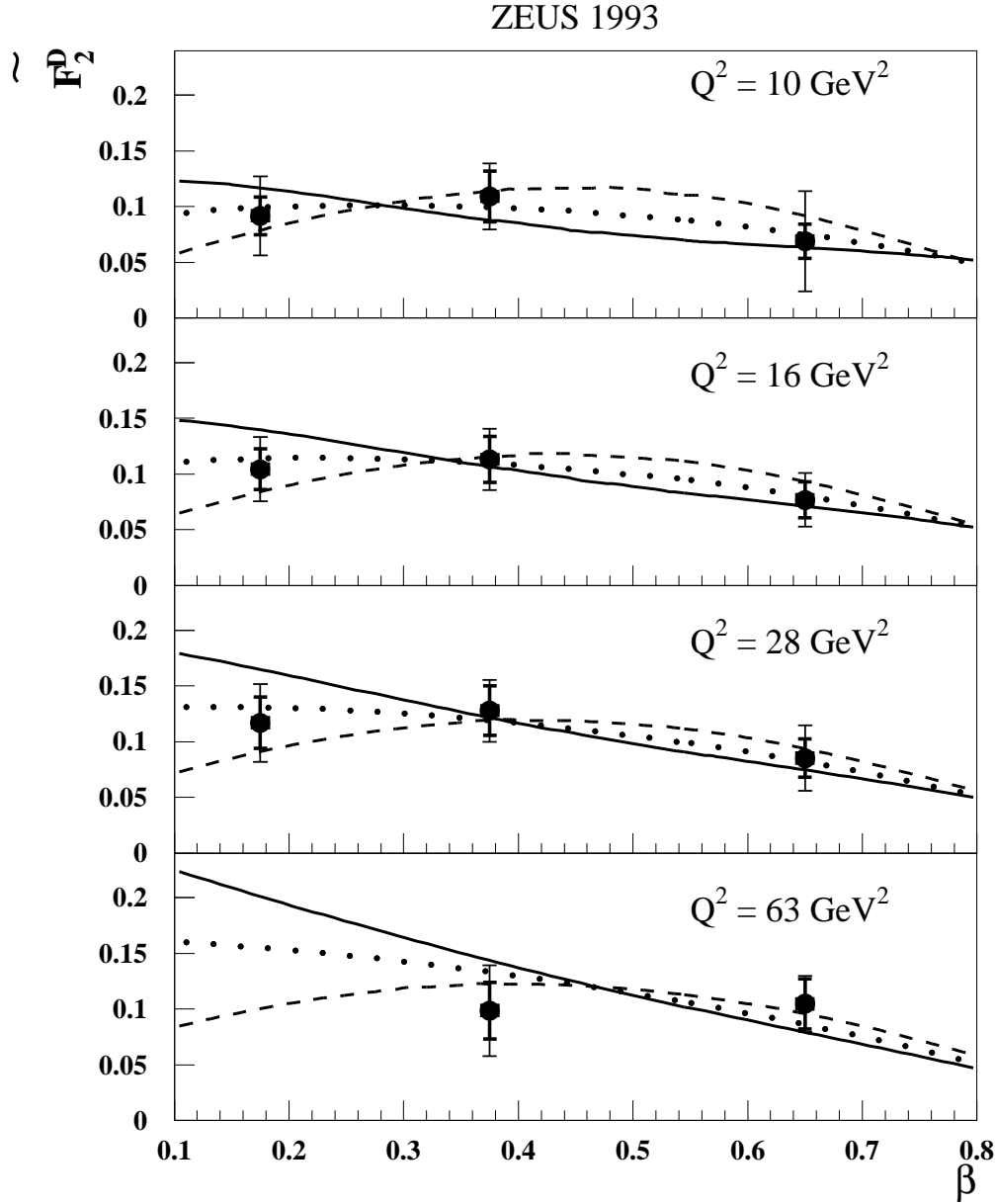}}
\caption{\label{figdra7}{ Measurements of $\tilde{F}_2^D(\beta,Q^2)$ in
 DIS \cite{zelrgd95} as a function of $\beta$ for fixed values of $\q2$
 compared to the results of the QCD fits: ``hard quark + hard gluon'' 
 (solid lines), ``hard quark + leading gluon'' (dashed lines) and 
 ``hard quark + (hard \& leading) gluon'' (dotted lines) parametrisations. 
 The measurements contain an estimated $(31\pm 13)\%$ contribution from double
 dissociation and the results of the QCD fits have been scaled up to
 take into account this contribution.}}
\end{figure}

 The fits were performed simultaneously to the 1993 ZEUS measurements of
$\tilde{F}_2^D(\beta,Q^2)$ (Figure~\ref{figdra7}) and the measured cross 
sections $\seta$ (Figure~\ref{figdra2}) and $\sbeta$ (Figure~\ref{figdra6})
in diffractive photoproduction. These cross sections in diffractive 
photoproduction are those which are most sensitive to the shape of the pomeron
parton densities. The following functional forms for the momentum-weighted
parton densities in the pomeron at $\mu^2_0$ have been used in the fits:
\begin{itemize}
\item hard quark + hard gluon:
  $$ \beta f_{u/\Pma}(\beta,\mu^2_0)= a_1 \beta (1-\beta) \; \; , \; \;
   \beta f_{g/\Pma}(\beta,\mu^2_0)= b_1 \beta (1-\beta); $$
\item hard quark + leading gluon:
  $$ \beta f_{u/\Pma}(\beta,\mu^2_0)= a_2 \beta (1-\beta) \; \; , \; \;
   \beta f_{g/\Pma}(\beta,\mu^2_0)= b_2 \beta^8 (1-\beta)^{0.3},$$
\item hard quark + (hard \& leading) gluon:
  $$ \beta f_{u/\Pma}(\beta,\mu^2_0)= a_3 \beta (1-\beta) \; \; , \; \;
   \beta f_{g/\Pma}(\beta,\mu^2_0)= b_3 \beta (1-\beta)+
   c_3 \beta^8 (1-\beta)^{0.3},$$
\end{itemize}
where the values of $a_i$, $b_i$ and $c_i$ are determined from the fits. 
Note that all these functional forms are constrained to be zero at $\beta=1$.
The exact shape of the parton distributions are only weakly constrained for
$\beta\gsim 0.8$.

 The fraction of the pomeron momentum carried by partons is defined as
 $$\Sigma_{\Pma}^{NLO} \equiv \int_0^1 d\beta \ \beta \ 
   [f_{g/\Pma}(\beta,\mu^2)+\sum_{j} f_{q_j/\Pma}(\beta,\mu^2)],$$
where the sum runs over all quark flavours which contribute at the given value
of $\mu^2$. Since the pomeron is not a particle it is unclear whether or not
the momentum sum rule ($\Sigma_{\Pma}^{NLO}=1$) should be satisfied.
Therefore, $\Sigma_{\Pma}^{NLO}$ has been left unconstrained in the fits.

 The functional form of the ``leading gluon'' distribution is similar to one
of the parametrisations extracted by the H1 Collaboration from the observed
scaling violations of the diffractive structure function \cite{h1f2d97}.
Parametrisations of the parton densities in which the pomeron is assumed to
be made exclusively of quarks have been disfavoured by previous 
measurements \cite{h1f2d97,zegpom95} and are not considered here. Likewise,
a parametrisation of the gluon density with a soft spectrum has been
ruled out \cite{h1f2d97,zegpom95,h1gp95}.

 The fitted values of the parameters are shown in Table~\ref{tabfit}. These
values have been obtained by subtracting from the data an estimated
$(31\pm 13)\%$ contribution due to double dissociation. In the figures, the
results of the fits have been scaled up to take into account this 
contribution. The results of the three different fits are compared to the
measurements used for the fits in Figures~\ref{figdra2}, \ref{figdra6} and
\ref{figdra7}. The calculations based on these fits, which include a substantial
hard momentum component of gluons at $\mu^2_0$ in the pomeron, give a
reasonable description of the shape and normalisation of the measurements.
The measured $\sbeta$ rises as $\betao$ increases while the calculations
fall at $\betao \gsim 0.8$ for the parton distributions chosen. This comparison
indicates that there is a sizeable contribution of gluons at large $\beta$.

\begin{figure}
\epsfysize=18cm
\centerline{\epsffile{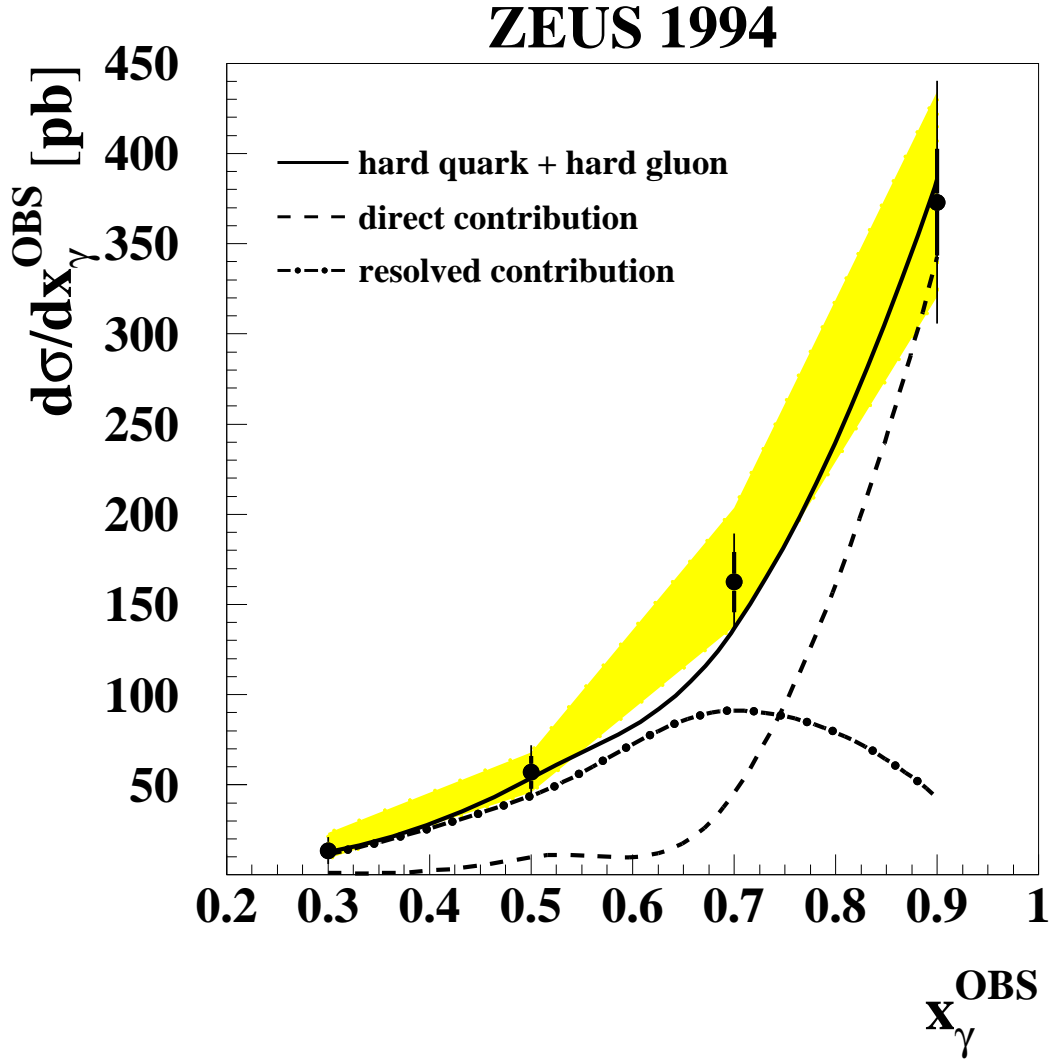}}
\vspace{-1.5cm}
\caption{\label{figdra5}{ Measured dijet cross section $\sxg$ integrated
 over $\etjet >6$~GeV and $\etar$. Other details as in Figure~\ref{figdra2}.
 For comparison, the calculations for the resolved (dot-dashed line), 
 direct (dashed line) and resolved plus direct processes (solid line) based
 on the QCD fit with the ``hard quark + hard gluon'' parametrisation
 (see text) are shown.}}
\end{figure}

 The results of the fits have also been compared to those dijet cross 
sections in diffractive photoproduction which were {\rm not} used in the fit
(Figures~\ref{figdra3} to \ref{figdra5}). The calculations
provide a good description of the measured $\set$, $\sw$ and $\sxg$. 
The sensitivity of the
measured $\sw$ to the value of $\alpha(0)$ in the pomeron trajectory is 
limited due to the cut on $\etjet$ used to define the jets and the
requirement on $\etam$.
These comparisons are consistent with the conclusion that there is
a large component of gluons with a hard momentum spectrum in the pomeron.
Moreover, the measured $\set$ is well described by the calculations,
indicating that the dynamics of dijet diffractive photoproduction is governed
by the matrix elements of perturbative QCD.

 The predicted contributions from direct and resolved processes to the 
measured $\sxg$, together with their sum, are shown in Figure~\ref{figdra5}. 
The sum of the contributions from resolved and direct processes gives a good 
description of the data. The shape of the contribution of either a purely 
direct-photon or a purely resolved-photon is not able to reproduce the data. 
A resolved-photon component is therefore needed in order to explain the shape
of the measured cross section for values of $\xg$ below 0.8. This observation
represents the first clear experimental evidence for the presence of both a
resolved- and a direct-photon component in diffractive dijet photoproduction.
Hard-scattering factorisation-breaking effects due to the resolved-photon
component might be suppressed by the observed dominance of the direct-photon
contribution to diffractive dijet photoproduction in the selected region of
phase space.

 The fraction of the pomeron momentum carried by partons which is due to
gluons, 
 $$c_g^{NLO}(\mu^2) \equiv \frac{1}{\Sigma_{\Pma}^{NLO}}
       \int_0^1 d\beta \ \beta \ f_{g/\Pma}(\beta,\mu^2),$$ 
depends upon the scale at which the parton content of the pomeron is probed
and has been computed for each of the fits. The determination of 
$c_g^{NLO}(\mu^2)$ is affected by the following uncertainties:
\begin{itemize}
 \item The statistical and systematic uncertainties of the measurements
       used in the fits, which are the dominant sources of uncertainty.
 \item The uncertainty ($\pm 30$\%) on the POMPYT calculations due to 
       higher-order QCD corrections.
 \item The uncertainty on the pomeron trajectory in the DL pomeron flux
       factor. The effect on $c_g^{NLO}(\mu^2)$ has been estimated by 
       changing $\alpha(0)$ from 1.085 to 1.15.
\end{itemize}
The central value of $c_g^{NLO}(4$~GeV$^2)$ varies
between 0.75 and 0.90 depending on the parametrisation and is given in
Table~\ref{tabfit}. Taking into account all the uncertainties of the three
fits, the resulting range for $c_g^{NLO}$ is 
$0.64< c_g^{NLO}(4$~GeV$^2) < 0.94$. 

\begin{figure}
\epsfysize=18cm
\centerline{\epsffile{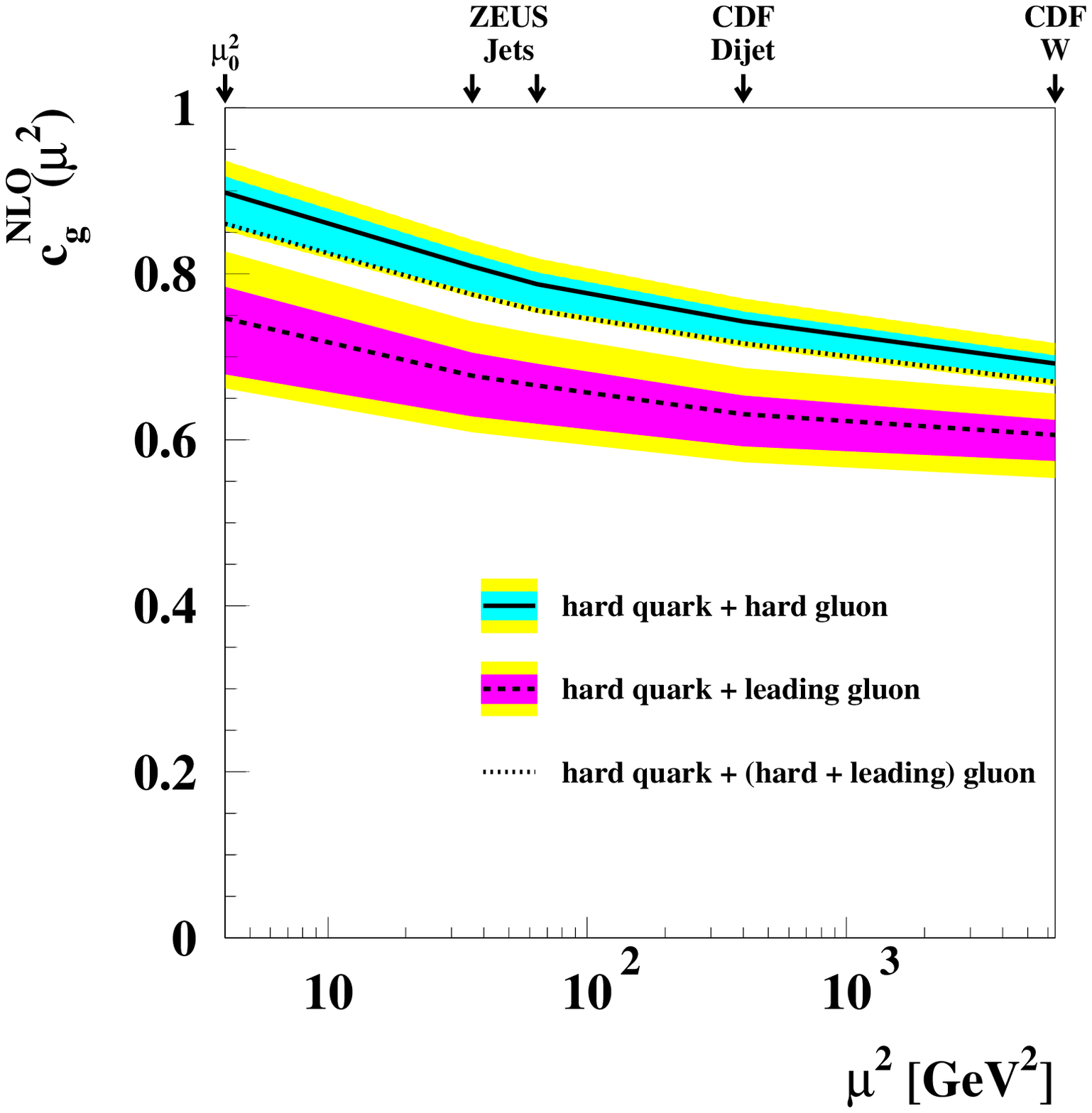}}
\caption{\label{figdra8}{ The values of $c_g^{NLO}(\mu^2)$ as a function
 of $\mu^2$ for each of the QCD fits described in the text. For the
 ``hard quark + hard gluon'' (upper band) and ``hard quark + leading gluon''
 (lower band) parametrisations, the central values (indicated by the lines),
 the statistical and systematic uncertainties $-$not associated with the
 absolute energy scale of the jets$-$ added in quadrature (light-shaded bands)
 and the uncertainty due to the absolute energy scale of the jets (dark-shaded
 bands) are shown. The theoretical uncertainties are included in the
 light-shaded bands and discussed in the text. For the ``hard quark + 
 (hard \& leading) gluon'' parametrisation, only the central values are shown
 (dotted line). The $\mu^2$ values at which $c_g^{NLO}(\mu^2)$ has been
 computed are indicated by the arrows.}}
\end{figure}

 The fraction $c_g^{NLO}(\mu^2)$ has also been computed at the $\mu^2$ values
probed\footnote{For the non-DIS measurements referred to here the $\mu^2$
value is not uniquely defined and $c_g^{NLO}$ has been computed at either
$(E_T^{jet})^2$ or the square of the $W$-boson mass depending on the
specific final state.} by other measurements and is shown in
Figure~\ref{figdra8}: 
a)~$\mu^2=(\etjet)^2=(6$~GeV$)^2=36$~GeV$^2$
 in dijet diffractive photoproduction (this analysis);
b)~$\mu^2=(\etjet)^2=(8$~GeV$)^2=64$~GeV$^2$ in inclusive jet diffractive 
 photoproduction \cite{zegpom95};
c)~$\mu^2=(\etjet)^2=(20$~GeV$)^2=400$~GeV$^2$ in dijet diffractive production
 in $\bar{p}p$ \cite{dcdf}; and 
d)~$\mu^2=M_W^2=6400$~GeV$^2$ in diffractive $W$-boson production in 
 $\bar{p}p$ \cite{wcdf}.

 The result for $c_g^{NLO}$ is consistent with the previous determination by 
ZEUS \cite{zegpom95} though now with an improved method and accuracy. The 
extrapolation of this result to the $\mu^2$ values probed by the measurements
in $\bar{p}p$ collisions is also consistent with the estimate of 
$(70\pm20)$\% given by the CDF Collaboration \cite{dcdf}. However, care must
be taken in these comparisons since the result presented here has been made
using pomeron parton densities which depend upon the scale according to
the DGLAP equations while those in \cite{zegpom95} and \cite{dcdf} 
neglected any scale dependence. The use of different procedures does not have
a significant effect on the extracted values of $c_g$, but affects
substantially the values of $\Sigma_{\Pma}$. 

 We conclude that for the selected region of phase space, which includes the 
$\etam$ requirement, and within our experimental uncertainties it is
possible to reproduce the measurements of the diffractive structure function 
\cite{zelrgd95} and of the dijet cross sections in diffractive
photoproduction with the Ingelman-Schlein model \cite{ingsch}. In this model,
which assumes Regge factorisation, the parton densities of the pomeron
evolve according to the DGLAP equations. The dominance of the direct process
for the measured region of phase space could be limiting the observation of
hard-scattering factorisation-breaking effects due to the resolved-photon
component. The data require the fraction of the pomeron momentum carried by
partons which is due to gluons to lie in the range
$0.64< c_g^{NLO}(4$~GeV$^2) < 0.94$.

\section{\bf Summary and conclusions}

 Measurements of the differential cross sections for dijet photoproduction 
with a large rapidity gap in $e^+p$ collisions at a centre-of-mass energy of 
300~GeV have been presented. The $e^+p$ dijet cross sections refer to jets at 
the hadron level with a cone radius of one unit in the $\etaphi$ plane. They
are given in the kinematic region defined by $\q2 \leq 4$ \g2\  (with a
median $\q2\approx 10^{-3}$~GeV$^2$) and \wr\ with the most-forward-going
hadron at $\etam < 1.8$. These cross sections have been measured as a
function of $\etajet$, $\etjet$ and $W$ for dijet production with
$\etjet >6$~GeV and $\etar$.

 The measured cross section $\sxg$ as a function of $\xg$, the fraction of
the photon momentum participating in the production of the two jets with
highest $\etjet$, peaks at $\xg \sim 1$ with a pronounced tail to lower
values. This result is clear evidence for resolved- and direct-photon
components in diffractive dijet photoproduction. 

 A measurement of the cross section for diffractive dijet photoproduction
as a function of $\betao$, the fraction of the pomeron momentum 
participating in the production of the two jets with highest $\etjet$, has
been presented. For the selected region of phase space the measured cross 
section $\sbeta$ increases as $\betao$ increases. This result shows that
there is a sizeable contribution to dijet production from those events in
which a large fraction of the pomeron momentum participates in the hard
scattering.

 A QCD analysis of the measurements of the diffractive structure function 
in DIS \cite{zelrgd95} and of the measured cross sections $\seta$ and
$\sbeta$ presented here has been performed. The pomeron is assumed to have
hadron-like partonic structure in the form of parton densities which evolve
according to the DGLAP equations. It is possible to reproduce both sets of 
measurements when a substantial hard momentum component of gluons in the
pomeron at the initial scale of $2$~GeV is included. 
The data require the fraction of the pomeron momentum carried by partons which
is due to gluons to lie in the range $0.64< c_g^{NLO}(4$~GeV$^2) < 0.94$.

\vspace{0.5cm}
\noindent {\Large\bf Acknowledgements}
\vspace{0.3cm}

 The strong support and encouragement of the DESY Directorate have been 
invaluable. The experiment was made possible by the inventiveness and the 
diligent efforts of the HERA machine group.  The design, construction and 
installation of the ZEUS detector have been made possible by the
ingenuity and dedicated efforts of many people from inside DESY and
from the home institutes who are not listed as authors. Their 
contributions are acknowledged with great appreciation. We would like to
thank L. Alvero and J.C. Collins for valuable discussions.


%
%

\newpage
\clearpage
\begin{table}
\begin{center}
\begin{tabular}[t]{|c|c||c@{~$~\pm~$}c@{~$~\pm~$}c|c|c|}\hline
$\etajet$-range  & Bin  & ${d\sigma}/{d\etajet}$
& stat.& syst.& syst. $\etjet$-scale & non-diff. subtr.\\
 & centre & \multicolumn{3}{c|}{[pb]}& [pb] & [pb] \\ 
\hline \hline
(-1.5,-1) &-1.25 &  75 & 8 & 13 &(+25,-17)& 10 \\ \hline
(-1,-0.5) &-0.75 & 118 &10 & 22 &(+30,-22)& 24 \\ \hline
(-0.5,0)  &-0.25 & 141 &11 & 21 &(+29,-23)& 31 \\ \hline
(0,0.5)   & 0.25 & 114 & 9 & 20 &(+25,-18)& 35 \\ \hline
(0.5,1)   & 0.75 &  70 & 7 & 13 &(+17,-13)& 24 \\ \hline \hline
$\etjet$-range & weighted mean & ${d\sigma}/{d\etjet}$
& stat.& syst.& syst. $\etjet$-scale &non-diff. subtr.\\
 {[GeV]} & [GeV] & \multicolumn{3}{c|}{[pb/GeV]}& 
[pb/GeV] & [pb/GeV]\\ \hline \hline
(6,8)  &  6.9 & 97   & 5   & 16  &(+22,-18)  & 17  \\ \hline
(8,10) &  8.8 & 25.2 & 1.7 &  3.7&(+6.2,-4.0)&  9.0 \\ \hline
(10,12)& 10.7 &  7.4 & 1.0 &  1.9&(+2.1,-1.5)&  3.1 \\ \hline
(12,14)& 12.8 &  2.1 & 0.5 &  0.7&(+0.5,-0.5)&  1.3 \\ \hline \hline
$W$-range & Bin centre & ${d \sigma}/{d W}$
& stat.& syst.& syst. $\etjet$-scale &non-diff. subtr.\\
 {[GeV]} & [GeV] & \multicolumn{3}{c|}{[pb/GeV] }& 
[pb/GeV]&[pb/GeV]\\ \hline \hline
(134,170) & 152 & 0.43 & 0.07 & 0.13&(+0.16,-0.10)& 0.11 \\ \hline
(170,206) & 188 & 0.98 & 0.12 & 0.15&(+0.24,-0.18)& 0.19 \\ \hline
(206,241) & 223 & 1.02 & 0.12 & 0.20&(+0.23,-0.17)& 0.27 \\ \hline
(241,277) & 259 & 1.14 & 0.09 & 0.22&(+0.25,-0.19)& 0.29 \\ \hline \hline
$\xg$-range  & Bin & ${d\sigma}/{d\xg}$
& stat.& syst.& syst. $\etjet$-scale &non-diff. subtr.\\
 & centre & \multicolumn{3}{c|}{[pb]}&[pb]&[pb]\\ \hline \hline
(0.2,0.4) & 0.3 &  13.6 &  3.6 &  7.5&( +9.8, -4.6)&  8.8\\ \hline
(0.4,0.6) & 0.5 &  57   &  9   & 15  &(+11  ,-11)  & 35  \\ \hline
(0.6,0.8) & 0.7 & 162   & 17   & 27  &(+41  ,-25)  & 46  \\ \hline
(0.8,1.0) & 0.9 & 373   & 30   & 67  &(+62  ,-52)  & 64  \\ \hline \hline
$\betao$-range  & Bin & ${d\sigma}/{d\betao}$
& stat.& syst.& syst. $\etjet$-scale &non-diff. subtr.\\
 & centre & \multicolumn{3}{c|}{[pb]}& [pb] &[pb]\\ \hline \hline
(0.4,0.55) & 0.48& 115 & 20 & 26&(+27,-20)&  6 \\ \hline
(0.55,0.7) & 0.63& 128 & 20 & 28&(+36,-26)& 22 \\ \hline
(0.7,0.85) & 0.78& 279 & 29 & 56&(+59,-51)& 54 \\ \hline
(0.85,1.0) & 0.93& 330 & 35 & 78&(+58,-40)& 73 \\ \hline
\end{tabular}
\end{center}
\caption{\label{tabsec1}{ Measured cross sections 
 in the kinematic region defined by $\q2\leq 4$ \g2\ and \mbox{\wr}\ 
 with the most-forward-going hadron at $\etam < 1.8$. 
 The contribution from non-diffractive processes, which is given in the last
 column, has been subtracted. The measurements contain an estimated
 $(31\pm 13)\%$ contribution from double dissociation.
 The statistical and systematic uncertainties $-$not associated with the 
 absolute energy scale of the jets$-$ are also indicated. 
 The systematic uncertainties associated to the absolute energy scale of the 
 jets are quoted separately. The overall normalisation uncertainty of 1.5\%
 from the luminosity determination is not included.}} 
\end{table}

\newpage
\clearpage
\begin{table}
\begin{center}
\begin{tabular}[t]{|c|c||c|c|c|c|c|c|}\hline
$i$  & Parametrisation & $a_i$ & $b_i$ & $c_i$ & $c_g^{NLO}(4$~GeV$^2)$ & 
$\Sigma_{\Pma}^{NLO}$ & $\chi^2_{stat}$/dof \\ \hline \hline
1& hard quark + hard gluon             & 0.30 & 10.5 & -    & 0.90 &1.96& 49/18
   \\ \hline
2& hard quark + leading gluon          & 0.34 & 13.2 & -    & 0.75 &0.89& 53/18
   \\ \hline
3& hard quark + (hard \& leading) gluon& 0.32 & 5.74 & 6.31 & 0.86 &1.49& 33/17
   \\ \hline \hline
\end{tabular}
\end{center}
\caption{\label{tabfit}{ Fitted values of the parameters for each
 of the three fits discussed in the text. The values of 
 $c_g^{NLO}(4$~GeV$^2)$, $\Sigma_{\Pma}^{NLO}$ and 
 $\chi^2_{stat}$, and the number of degrees
 of freedom (dof) for each of the fits are also shown.}}
\end{table}
%
%
%
%

\newpage
\clearpage
\parskip 0mm

\newpage
\clearpage
\parskip 0mm

\newpage
\clearpage
\parskip 0mm

\newpage
\clearpage
\parskip 0mm

\newpage
\clearpage
\parskip 0mm

\newpage
\clearpage
\parskip 0mm

\newpage
\clearpage
\parskip 0mm

\newpage
\clearpage
\parskip 0mm


\begin{thebibliography}{99}
\bibitem{ingsch} G. Ingelman and P.E. Schlein, Phys. Lett. B152 (1985) 256.

\bibitem{berger} E.L. Berger et al., Nucl. Phys. B286 (1987) 704.

\bibitem{donlan} A. Donnachie and P.V. Landshoff, Nucl. Phys. B303 (1988) 634.

\bibitem{streng} K.H. Streng, Proc. of HERA Workshop, DESY (1987) 365.

\bibitem{nikzak} N.N. Nikolaev and B.G. Zakharov, Z. Phys. C53 (1992) 331.

\bibitem{donlan2} A. Donnachie and P.V. Landshoff, Phys. Lett. B285 (1992) 
  172.

\bibitem{cfs} J.C. Collins, L. Frankfurt and M. Strikman, Phys. Lett. B307 
  (1993) 161.

\bibitem{soper} A. Berera and D.E. Soper, Phys. Rev. D50 (1994) 4328.

\bibitem{ua8coll} UA8 Collab., A. Brandt et al., Phys. Lett. B211 (1988) 239, 
  B297 (1992) 417.

\bibitem{zelrgd93} ZEUS Collab., M. Derrick et al., Phys. Lett. B315 (1993) 
  481, Phys. Lett. B332 (1994) 228.

\bibitem{h1lrgd94} H1 Collab., T. Ahmed et al., Nucl. Phys. B429 (1994) 477.

\bibitem{h1lrgd95} H1 Collab., T. Ahmed et al., Phys. Lett. B348 (1995) 681.

\bibitem{zelrgd95} ZEUS Collab., M. Derrick et al., Z. Phys. C68 (1995) 
  569.

\bibitem{zef2d497} ZEUS Collab., J. Breitweg et al., Eur. Phys. J. C1 (1998)
  81.

\bibitem{h1f2d97} H1 Collab., C. Adloff et al., Z. Phys. C76 (1997) 613.

\bibitem{dglap} L.N. Lipatov, Sov. J. Nucl. 20 (1975) 95; V.N. Gribov and
  L.N. Lipatov, Sov. J. Nucl. Phys. 15 (1972) 438; G. Altarelli and
  G. Parisi, Nucl. Phys. B126 (1977) 298; Yu. L. Dokshitzer, Sov. Phys.
  JETP 46 (1977) 641.

\bibitem{zegpom95} ZEUS Collab., M. Derrick et al., Phys. Lett. B356 (1995) 
  129.

\bibitem{wcdf} CDF Collab., F. Abe et al., Phys. Rev. Lett. 78 (1997) 2698.

\bibitem{dcdf} CDF Collab., F. Abe et al., Phys. Rev. Lett. 79 (1997) 2636.

\bibitem{juan} L. Alvero, J.C. Collins, J. Terron and J. Whitmore,
  ``Diffractive Production of Jets and Weak Bosons, and Tests of
    Hard Scattering Factorization'', preprint CTEQ-701 (hep-ph/9701374).

\bibitem{goul} K. Goulianos, ``Factorization and Scaling in Hard
   Diffraction'', talk given at 5th International Workshop on Deep Inelastic
   Scattering and QCD (DIS 97), Chicago, IL, 14-18 April 1997 (hep-ph/9701374).

\bibitem{owens} J.F. Owens, Phys. Rev. D21 (1980) 54.

\bibitem{drees} W.J. Stirling and Z. Kunszt, Proceedings of the HERA Workshop 
 (1987) 331; M. Drees and F. Halzen, Phys. Rev. Lett. 61 (1988) 275;
 M. Drees and R.M. Godbole, Phys. Rev. D39 (1989) 169.

\bibitem{zenov93} ZEUS Collab., M. Derrick et al., Phys. Lett. B322 (1994)
 287.

\bibitem{zedij95} ZEUS Collab., M. Derrick et al., Phys. Lett. B348 (1995)
 665.

\bibitem{facthe} J.C. Collins, D.E. Soper and G. Sterman, Nucl. Phys. B261
  (1985) 104 and B308 (1988) 833; G.T. Bodwin, Phys. Rev. D31 (1985) 2616
  and D34 (1986) 3932.

\bibitem{collproof} J.C. Collins, ``Proof of Factorization for Diffractive
  Hard Scattering'', Preprint PSU/TH/189 (hep-ph/9709499) and
  references therein.

\bibitem{ingpry} G. Ingelman and K. Jansen-Prytz, Z. Phys. C58 (1993) 285.

\bibitem{sigtot} ZEUS Collab., M. Derrick et al., Phys. Lett. B293 (1992) 465.

\bibitem{status} The ZEUS Detector, Status Report (1993), DESY 1993.

\bibitem{pythia} H.-U. Bengtsson and T. Sj\"ostrand, Comp. Phys.
      Comm. 46 (1987) 43; T. Sj\"ostrand, Comp. Phys. Comm. 82 (1994) 74.

\bibitem{grv} M. Gl\"uck, E. Reya and A. Vogt, Phys. Rev. D46 (1992) 1973.

\bibitem{mrsa} A.D. Martin, W.J. Stirling and R.G. Roberts, Phys. Rev.
      D50 (1994) 6734.

\bibitem{lund} B. Andersson et al., Phys. Rep. 97 (1983) 31.

\bibitem{jetset} T. Sj\"ostrand, Comp. Phys. Comm. 39 (1986) 347;
  T. Sj\"ostrand and M. Bengtsson, Comp. Phys. Comm. 43 (1987) 367.

\bibitem{pompyt} P. Bruni and G. Ingelman, Proc. of the International 
  Europhysics Conference, edited by J. Carr and M. Perrotet, Marseille, France,
  July 1993 (Ed. Frontieres, Gif-sur-Yvette, 1994), p. 595.

\bibitem{gs} L.E. Gordon and J.K. Storrow, Z. Phys. C56 (1992) 307.

\bibitem{zeoct97} ZEUS Collab., J. Breitweg et al., Eur. Phys. J. C2 (1998) 61.

\bibitem{snow} J. Huth et al., Proc. of the 1990 DPF Summer Study on
  High Energy Physics, Snowmass, Colorado, edited by E.L. Berger
  (World Scientific, Singapore,1992) p. 134.

\bibitem{zeoct94} ZEUS Collab., M. Derrick et al., Phys. Lett. B342 (1995)
  417.

\bibitem{jacblo} Method proposed by F. Jacquet and A. Blondel in Proc.
  of the Study for an $ep$ Facility for Europe, U. Amaldi et al.,
  DESY 79/48 (1979) 377.

\bibitem{zemx97eps} ZEUS Collab., ``Measurement of the Diffractive Cross
  Section in DIS at HERA'', contributed paper N-638 to the 
  International Europhysics Conference on High Energy Physics, Jerusalem,
  Israel, 1997.

\bibitem{VMN639eps} ZEUS Collab., ``Exclusive Vector Meson production in DIS
  at HERA'', contributed paper N-639 to the 
  International Europhysics Conference on High Energy Physics, Jerusalem,
  Israel, 1997.

\bibitem{juanpu} J. Puga, Ph.D. Thesis, Universidad Aut\'onoma de Madrid (1998).

\bibitem{dij97} ZEUS Collab., J. Breitweg et al., Eur. Phys. J. C1 (1998) 109.

\bibitem{collins1} J.C. Collins et al., Phys. Rev. D51 (1995) 3182;
  see also \cite{juan}.

\bibitem{cteq} CTEQ Collab., R. Brock et al., Rev. Mod. Phys. 67 (1995) 157. 

\bibitem{h1gp95} H1 Collab., T. Ahmed et al., Nucl. Phys. B435 (1995) 3.

\end{thebibliography}
\end{document}